\documentclass{tufte-handout}

\usepackage{amsmath}
\usepackage[]{stmaryrd}
\usepackage[]{fdsymbol}

\usepackage{array}

\usepackage{graphicx}
\setkeys{Gin}{width=\linewidth,totalheight=\textheight,keepaspectratio}
\graphicspath{{graphics/}}

\title{A taxonomy of video lecture styles}
\author[choko]{Konstantinos Chorianopoulos\thanks{This document is a preprint copy by the author, please use the following official reference to cite it in your work: \emph{Chorianopoulos, K. (2018). A taxonomy of asynchronous instructional video styles. The International Review of Research in Open and Distributed Learning. 19(1)}}}
\date{15 January 2018}  

\usepackage{booktabs}

\usepackage{units}

\usepackage{fancyvrb}
\fvset{fontsize=\normalsize}

\usepackage{multicol}

\usepackage{lipsum}

\begin{document}

\maketitle

\begin{abstract}
\noindent Many educational organizations are employing instructional video in their pedagogy, but there is limited understanding of the possible presentation styles. In practice, the presentation style of video lectures ranges from a direct recording of classroom teaching with a stationary camera and screencasts with voice-over, up to highly elaborate video post-production. Previous work evaluated the effectiveness of several presentation styles, but there has not been any consistent taxonomy, which would have made comparisons and meta-analyses possible. In this article, we surveyed the research literature and we examined contemporary video-based courses, which have been produced by diverse educational organizations and teachers across various academic disciplines. We organized video lectures in two dimensions according to the level of human presence and according to the type of instructional media. In addition to organizing existing video lectures in a comprehensive way, the proposed taxonomy offers a design space that facilitates the choice of a suitable presentation style, as well as the preparation of new ones.
\end{abstract}


\begin{figure*}
\includegraphics[width=.17\linewidth]{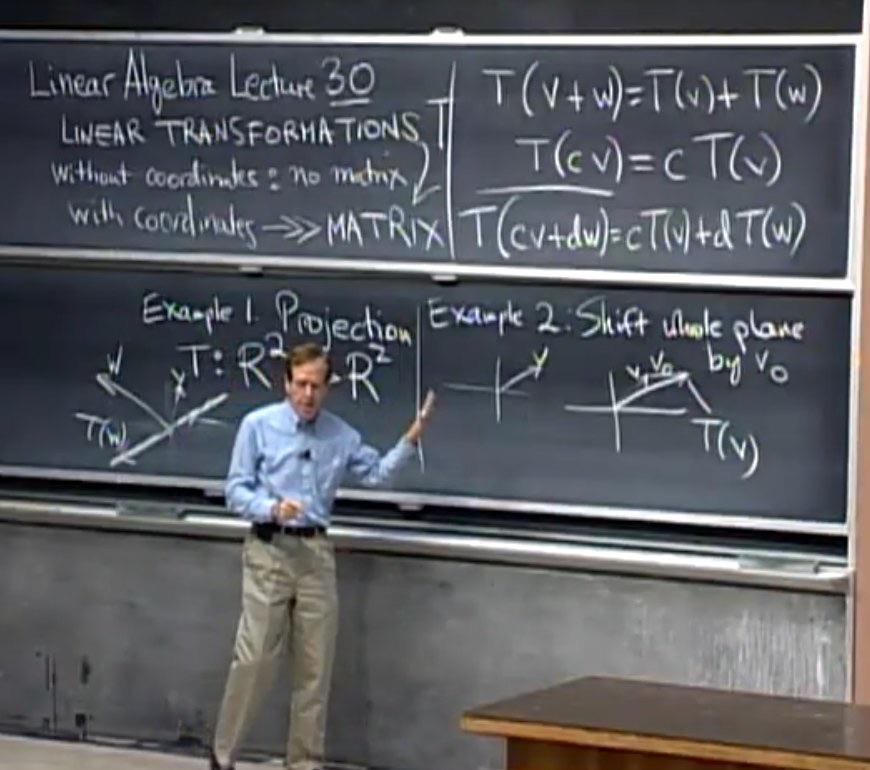}
\includegraphics[width=.2\linewidth]{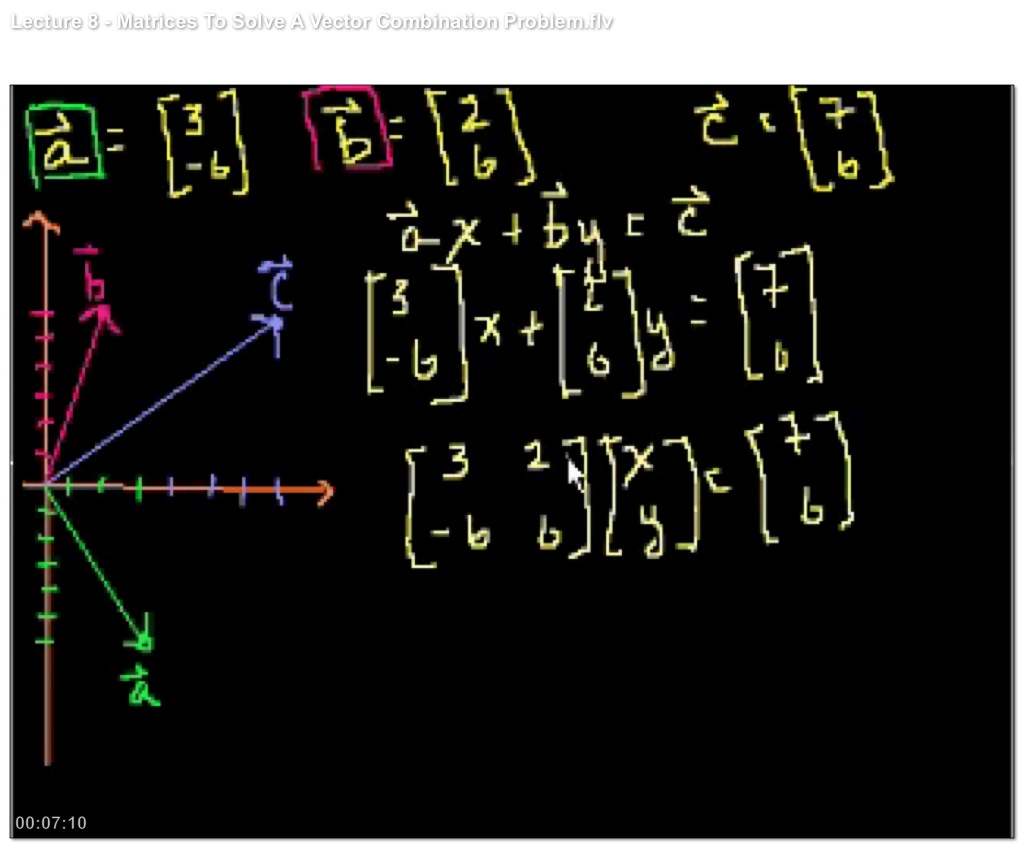}
\includegraphics[width=.3\linewidth]{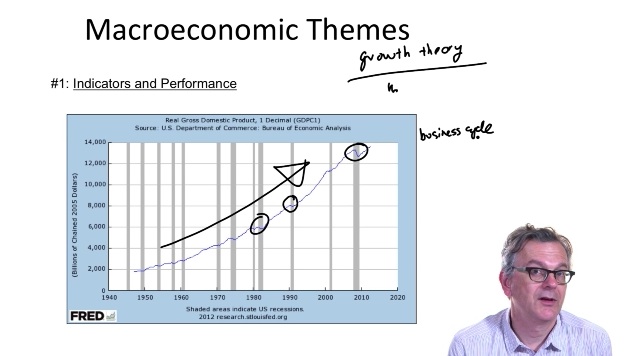}
\includegraphics[width=.25\linewidth]{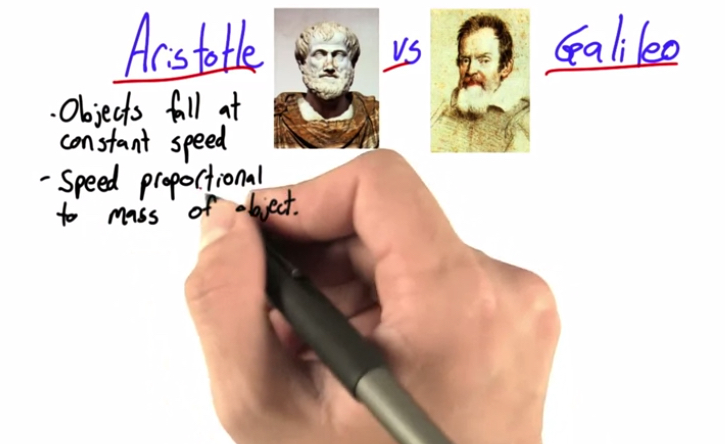}
\caption{Popular contemporary
video-based learning platforms provide widely different instructional video
formats, but there is limited understanding of the main production and
style factors that guide the selection of one over another: MIT
OpenCourseware, Khan Academy, Coursera, Udacity}
\end{figure*}

\newthought{Video lectures have been growing} and many individuals, organizations, and universities are employing them in various instructional frameworks, such as distance education and flipped classrooms. Alongside the wide availability of online video lectures on various topics, there is also a wide diversity of video production styles. In many cases, the same topic (e.g., statistics), is transferred to the instructional video format with very different presentation styles. Thus, every organization, or individual that produces instructional videos has to make an informed decision on the available video styles. Nevertheless, there is limited documentation on the main attributes of each instructional video style. Therefore, there is a need to organize instructional video styles in a simple way that facilitates the choice and the creation of novel styles.

In order to create a taxonomy of instructional video production styles, we need to identify the main distribution platforms, as well as representative examples. Besides the educational video repositories (e.g., YouTubeEdu, iTunesU, MIT Open Courseware, VideoLectures.net, TEDEd, etc), there is also a growing number of organizations that offer video-based learning, such as Coursera, Udacity, EdX, Khan Academy, FutureLearn, and Iversity. A survey of the available videos and of the state-of-the-art has revealed that the selection of a production style for a video lecture depends on the instructor's preference and feasibility, or on the organizational (platform) guidelines, rather than a structured theory. For example, all Udacity videos have the same presentation style. On the other hand, the videos on YouTubeEdu (a video lecture repository) have significant variability. In contemporary instructional video repositories, the most popular production style is the direct recording at the classroom, or at the teacher office with one camera.

There are several video platforms that facilitate established universities and instructors in preparing and sharing instructional video. In addition to the generic video platforms (e.g., YouTubeEdu, iTunesU), in the USA there are Coursera, EdX, and Udacity. In Europe, the main ones are FutureLearn (UK) and Iversity (EU). Notably, each one of the major video-based education providers seems to feel very confident about its approach to video presentation style and has a consistent style across the video archive and across different subjects and disciplines. For example, Shalman Khan, the founder of the Khan Academy has noted (Thompson 2011)\cite{thompson_2011}: "That way, it does not seem like I am up on a stage lecturing down at you. It is intimate, like we are both sitting at a table and we are working through something together, writing on a piece of paper."

On the other hand, the Coursera platform suggests a teacher-centered presentation style, which positions the teacher next to the slides, or at an over-imposed small window (picture-in-picture). Finally, Udacity takes the middle road and displays mostly the hand of the teacher, who writes and gestures on an interactive drawing board. Despite the major differences in the production styles, there are also some common patterns, such as the presence of humans and the use of complementary instructional media. In this article, we are exploring the main instructional video classification factors and their nuances.

In the rest of this article, we describe the main factors that affect the presentation style of video lectures. For this purpose, we have analyzed the state-of-the-art in the research literature and in the industry. The analysis of the research literature was based on extensive Google Scholar keyword searches (e.g., "instructional video", "video lecture", "MOOC") and the selection of a few recent articles (later than 2010) that have a very good number of citations per year (more than two yearly). The academic articles provided the theoretical groundwork for the taxonomy of the video lectures, which was then refined by performing a review of presentation styles in educational video repositories. The review of presentation styles was focused on a few major video platforms (YouTubeEdu, iTunesU, Coursera, Udacity, Khan Academy) and was aimed at collecting representative examples of different video styles, without regarding their actual popularity, or other aspects of production quality.

\section*{Survey of instructional video styles in scholarly  publications}

Researchers have recognized that different video production styles might have different learning effects. In the largest study of video lecture presentation styles, Guo et al. (2014)\cite{guo2014video} have identified six basic types of video production style: 1) classroom lecture with instructor on the blackboard, 2) talking head of instructor at desk, 3) digital drawing board (Khan-style), 4) slide presentation, 5) studio without audience, and 6) computer coding session. Some notable findings include that students prefer short videos, slides should include a talking head, the Khan drawing style is more engaging than slides or coding sessions, and the direct classroom recording does not work well online. Nevertheless, in their study, all the courses were from the same platform (EdX) and all the courses are from science and engineering. Although previous empirical research has provided many insights about several video styles, the aggregate results are not comparable because they do not have a common ground with regard to the typology of video production styles. For example, there are many studies that have examined screencasts, but ilioudi et al. (2013)\cite{ilioudi2013investigating} mentions Khan-style, which is technically a particular type of screencast that records the pen-tip of the presenter on a digital drawing board. Therefore, there is a semantics issue with regard to the unit of analysis that might reduce the understanding, comparison, and extension of previous works.

In order to identify existing instruction video styles and resolve any possible terminology ambiguities, we have organized previous works according to two recurring themes: instructional media (e.g., slides, animation, type) and human embodiment (e.g., social presence, animated human, talking-head). Indeed, Santos-Espino et al. (2016)\cite{santos2016speakers} examined the instructional video styles in contemporary MOOC platforms and classified them into two main categories: speaker-centric and board-centric. Although there are few instructional videos that employ just one style, they found that courses in humanities and arts emphasize the former, while science and engineering ones emphasize the latter. Social and life sciences employ a balanced approach between the speaker-centric and board-centric styles. Although they emphasize the use of the terms "speaker-centric" and "board-centric", we could generalize these two terms to the broader notions "human embodiment" and "instructional media" respectively, which are more common in the learning sciences. Moreover, it is worth considering the nuances along the two dimensions: human-embodiment and instructional media.

There are many studies that define a low-level unit of analysis that regards very detailed aspects of instructional media. Sugar et al. (2010)\cite{sugar2010examining} have provided an analysis of instructional videos, which are based on the screencasting style (recording of the screen). They found that there are two types of screen movement: static or dynamic (follows the cursor). They also found that there are two types of narrative: explicit that describes the exact actions on the screen and implicit, which describes the type of activity on the screen. Swarts (2012)\cite{swarts2012new} examined screencasting videos with a focus on multimedia software courses and provided guidelines for the production of good video tutorials that belong to the screencasting presentation style. Video tutorials that explain the use of particular features of computer software are a very popular category and many computer users prefer to watch a demonstration than reading a paper manual. As a matter of fact, the popularity of these video tutorials has also made popular the screencasting style of video instruction. Cross et al. (2013)\cite{cross2013typerighting} emphasized the use of digital writing on instructional video and compared the use of handwriting to typefaces. They found that learners preferred handwriting, but they considered more legible the typefaces, so they proposed a middle of the road approach that fades hand-writing into a typeface as soon as a word is complete. ten Hove and van der Meij (2015)\cite{ten2015like} analyzed the popularity of instructional videos on YouTube. Although "popularity" is not always correlated with effective pedagogy, it is indicative of contemporary learner expectations. They found that popular instructional videos shared some common characteristics such as fast-pace, text highlights, static images and animations, and high-definition production. Their analysis is the first that correlates particular production elements to effectiveness, but those elements are related mostly to planning and post-production, rather than presentation style. The above studies provide many insights into the design of instructional media in the context of video lectures, but there is no coherent framework, besides putting them in the same category.

Another major category of studies regards the presence as well as the type of human embodiment in the instructional video. Lyons et al. (2012)\cite{lyons2012video} performed a longitudinal study (13 weeks), which compared the use of video lectures with (or without) a video of the instructor at the top left of the screen. They found that students considered the social presence of the instructor in the video to offer more learning. Ilioudi et al (2013)\cite{ilioudi2013investigating} compared the Khan-style to the classroom recording and found that there were no major differences in preference, or learning performance between the two conditions. Chen and Wu (2015)\cite{chen2015effects} compared three popular instructional video styles: 1) direct classroom recording, 2) studio recorded video lectures with the video of the instructor superimposed to the slides, and 3) office recording of the instructor video next to the slides (voice-over type). Although the latter style includes the presence of the instructor, they refer to it as "voice-over type", because the slides and the voice are the main elements. Kizilcec et al. (2015)\cite{kizilcec2015instructor} compared the constant inclusion of a talking head in the video of a slide presentation to one with a moderate presence of the face of the presenter. They found that there were no significant differences in terms of learning performance, but it seems that some students just prefer the presence of the instructor's face. Li et al. (2016)\cite{li2016social} examined the acceptance of a virtual avatar in place of the instructor's talking head. They compare multiple alternatives in the place of the human instructional video, such as animated human, animated robot, and real robot. They found that learners preferred the real or animated human condition to the real or animated robot, but the recall rates were mixed across conditions and genders. Mayer and DaPra (2012)\cite{mayer2012embodiment} have found that learners prefer a human-like (e.g., voice, eye-contact, gestures) animated character. In particular, learners preferred real human voice rather than computerized voice. The above studies provide many insights about the presence and types of human embodiment in the context of video lectures, but there is no coherent framework, besides putting them in the same category.

\begin{table}
\centering
\begin{tabular}{>{\raggedleft}p{0.5\columnwidth}l}
    \toprule
    Style & Reference \\
    \midrule
    screencasting, screen movement, narration & Sugar et al. (2010) \\
    screencasting & Swarts (2012) \\
    animated human & Mayer and DaPra (2012) \\
    social presence & Lyons et al. (2012) \\
    screencasting, khan-style, handwriting, typeface & Cross et al. (2013) \\
    khan-style, classroom & Ilioudi et al. (2013) \\
    khan-style, classroom, studio, office-desk, code, slides & Guo et al. (2014) \\
    classroom, voice-over, picture-in-picture & Chen and Wu (2015) \\
    Static and dynamic pictures, text & ten Hove and van der Meij (2015 \\
    slides, talking head & Kizilcec et al. (2015) \\
    talking-head, robot, animated human, animated robot & Li et al. (2016) \\
    speaker-centric, board-centric & Santos-Espino et al. (2016) \\
    \bottomrule
\end{tabular}
\setfloatalignment{b}
\caption{Previous research has organized instructional video with overlapping or ambiguous tags and
categories, but there is no unifying framework, which might be confusing
in assessing contribution in further research}
\end{table}

In summary, previous research on the production style of video lectures has provided evaluations of particular presentation styles, but it has not done so in the context of a consistent taxonomy. The organization of previous research in a table that highlights the main classification factors of instructional video styles facilitates some observations. Firstly, there are differences between the formats tested by previous research. Notably, there are overlaps and ambiguities due to the lack of a common framework and terminology. For example, some works have emphasized low-level elements (e.g., typeface), while other works have employed different terms for similar concepts (e.g., talking head, voice-over). In this way, the production style of a video lecture has been classified in many different, or overlapping categories, which makes it difficult to compare across studies, or to perform meta-analyses. Ideally, a coherent taxonomy would be inclusive of all existing styles and should facilitate the informed choice for the production of a new instructional video. Moreover, a future-proof taxonomy should also hold the predictive attribute, which facilitates the definition of new production styles that do not yet exist.

In the next section, we examine contemporary instructional video styles, as found on online learning systems and educational video repositories, in order to identify nuances across the two dimensions of the proposed classification scheme: human embodiment and instructional media.

\section{Taxonomy of instructional video styles}

\begin{figure*}
\includegraphics[width=.17\linewidth]{figures/instructor-blackboard.jpg}
\includegraphics[width=.23\linewidth]{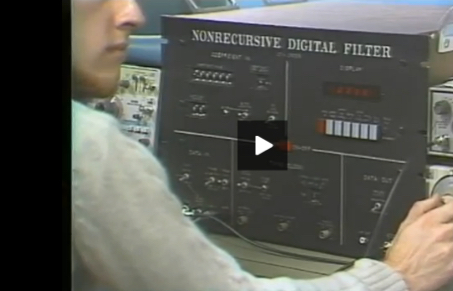}
\includegraphics[width=.23\linewidth]{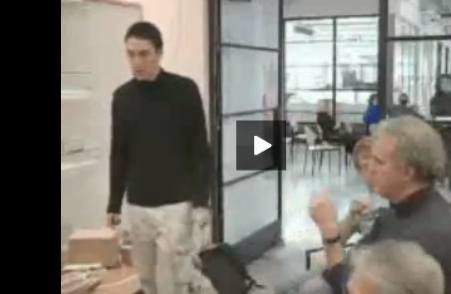}
\includegraphics[width=.27\linewidth]{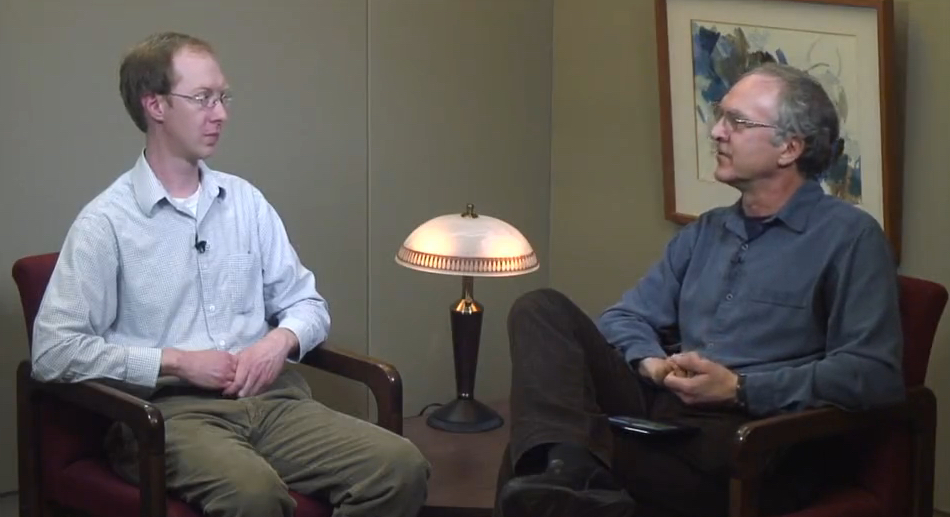}
\caption{The first quadrant includes styles that represent physical embodiment and physical boards: MIT OpenCourseware, iTunesU}
\end{figure*}

The availability of instructional video has been increasing since the early 2000’s, when broadband access from home became more affordable for more people. Initially, video lectures appeared on educational video repositories, such as YouTube and iTunes University. Next, video lectures spread quickly to specialized educational organizations, such as MIT Open Courseware, TEDed, Videolectures.net. Last, but not least, the instructional video format has become even more popular and refined within the Massive Open Online Courses (MOOCs), which have complemented video lectures with other popular e-learning elements, such as syllabus, e-books, assignments, discussion forums, wikis, and peer-grading. Although MOOCs are much more than just video lectures, the MOOC platforms have put a lot of effort in evolving the video lecture format. In this work, we focused on educational videos found on major educational video repositories (e.g., YouTubeEdu, iTunesU, MIT Open Courseware, VideoLectures.net, TEDEd) and platforms that offer video-based learning, such as Coursera, Udacity, EdX, Khan Academy, FutureLearn, and Iversity. Besides the classification of existing video lectures, the proposed taxonomy has put special emphasis on the granularity of the main factors (human-embodiment, instructional media) that define the classification of each presentation style.

\begin{figure*}
\includegraphics[width=.18\linewidth]{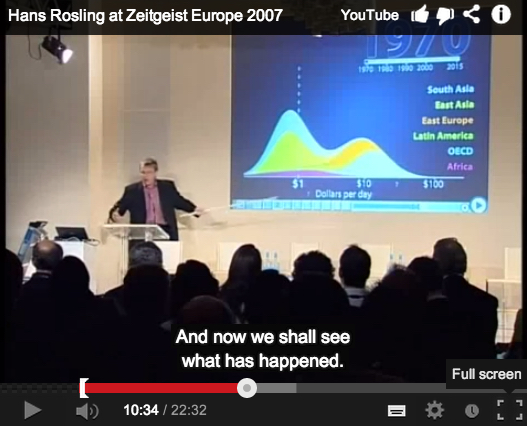}
\includegraphics[width=.19\linewidth]{figures/khan.jpeg}
\includegraphics[width=.25\linewidth]{figures/udacity.png}
\includegraphics[width=.25\linewidth]{figures/coursera-production-result.jpg}
\includegraphics[width=.22\linewidth]{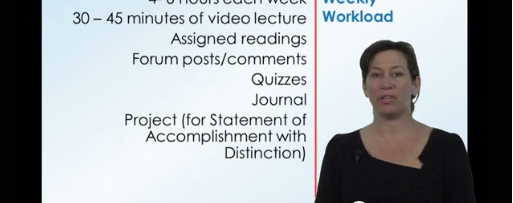}
\includegraphics[width=.21\linewidth]{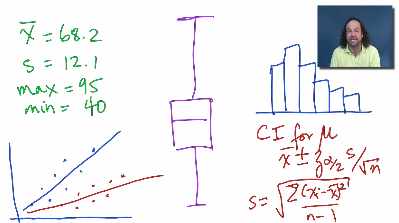}
\includegraphics[width=.18\linewidth]{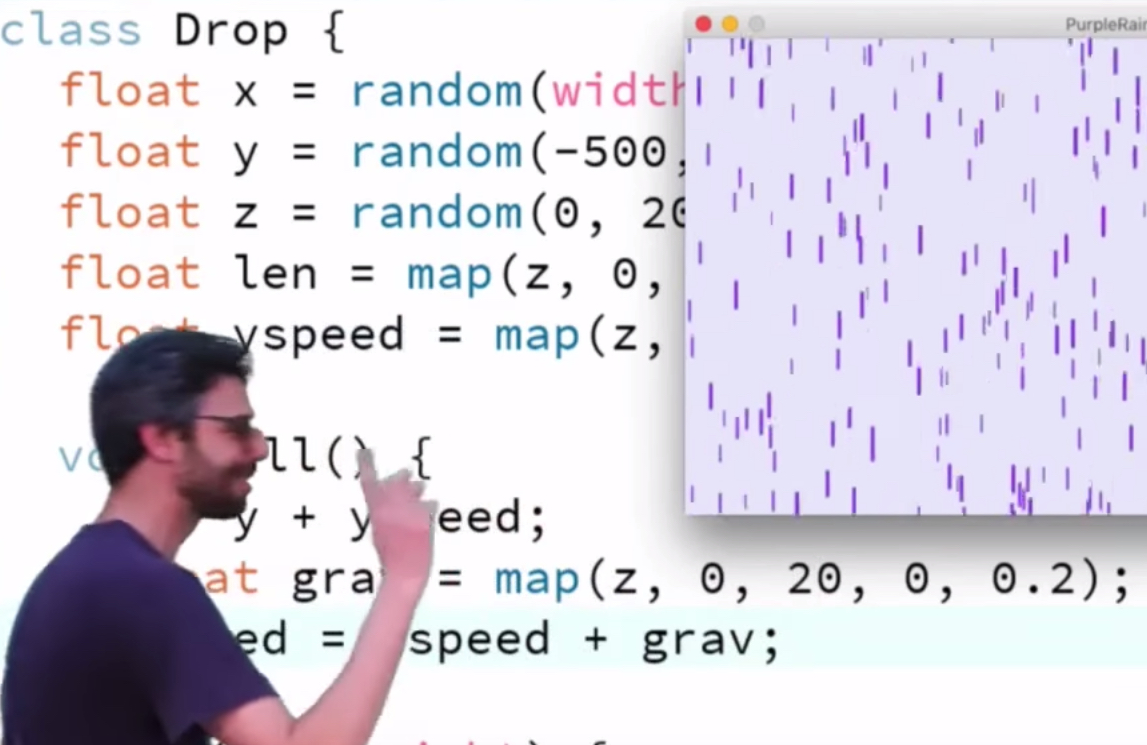}
\includegraphics[width=.16\linewidth]{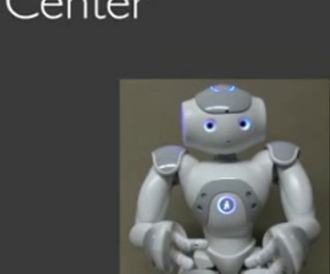}
\caption{The fourth and most popular quadrant includes styles that represent varying degrees of physical embodiment and strong emphasis on digital instructional media: TED, Khan Academy, Udacity, Coursera}
\end{figure*}

We performed a breadth-first random sampling of the available instructional video styles, in order to classify them in a Table (Table 1) according to two factors: human embodiment and instructional media. We explored hundreds of instructional videos, as many as required before we could not find any significantly different ones. The starting point of our taxonomy is the work of Santos-Espino et al. (2016)\cite{santos2016speakers}, which we extend by generalizing and by providing a more nuanced spectrum of nominal values along the two dimensions. In the proposed taxonomy, there are two dimensions that determine the presentation style: 1) human embodiment, 2) instructional media. For each one of the two dimensions there are multiple nominal values that range from the digital (or artificial) to the physical. Firstly, we examined the main presentation styles in order to assign nominal values to the main attributes, and next, we assigned each style on the cartesian table with the respective index symbol, in order to make the scatter-plot visualization more legible.

The proposed classification is qualitative and aims to reveal the existing presentation styles. It does not provide any information with the regard to popularity, or with regard to learning effectiveness, or suitability to a particular pedagogy, which are left to further work. In particular, the classification factors are nominal rather that quantitative, so the classification is not meant to be absolute about the particular nominal values. It is meant to be exact about clearly defining presentation styles before measuring them quantitatively. Both classification factors (human-embodiment, instructional media) have the same limits, which are fully digital and fully physical. Thus, the human-embodiment factor has been organized with nominal values the reflect a gamut of human-presence. Similarly, the instructional media factor has been organized with nominal values that reflect a gamut of different instructional media. Then, the mapping of existing instructional video styles on a nominal scatter plot is a straightforward visualization from the classification table, according to the index keys in the first column.

\begin{marginfigure}
\includegraphics[width=.8\linewidth]{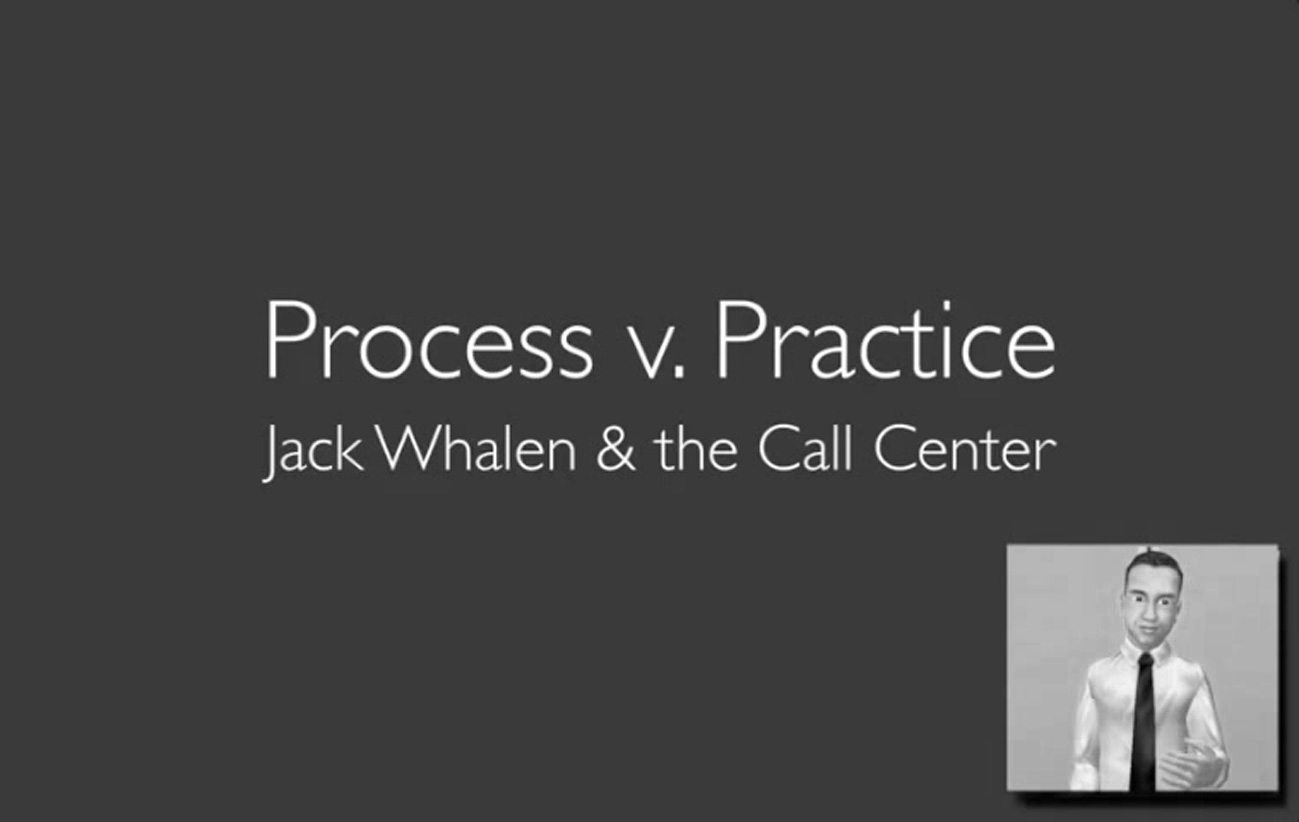}
\caption{The third quadrant includes styles that represent digital embodiment and digital boards}
\end{marginfigure}

\marginnote{Sources of video screenshots: TED, iTunesU, Khan Academy, Udacity, Coursera, Coursera, Coursera, MIT OpenCourseware, MIT OpenCourseWare, Coursera, YouTube, \citep{li2016social}}

In summary, video lectures can be organized in two dimensions: human embodiment and instructional media, which have several nominal values from digital to physical. Although the proposed taxonomy of presentation styles is just a snapshot of the current situation, the focus of the taxonomy is on the classification factors (human embodiment, instructional media) and the particular attributes (e.g., hand, face, slides, etc.), rather than the details of the production style. Therefore, the discussion that follows is based on those factors. In particular, the level of the human embodiment varies widely between video lectures from wide shots that include the audience heads to screen capturing of the tip of the pen. The type of instructional media is another classification factor that varies from slides and animations, to objects manipulated by the instructor. The proposed taxonomy should be useful in understanding the landscape of available options when planning to create a familiar instructional video, or when designing a novel presentation style.

\begin{table*}
\centering
\fontfamily{ppl}\selectfont
\begin{tabular}{p{0.11\columnwidth}p{0.1\columnwidth}p{0.11\columnwidth}p{0.13\columnwidth}p{0.15\columnwidth}p{0.15\columnwidth}}
\includegraphics{figures/audience.jpg} & \includegraphics{figures/instructor-blackboard.jpg} & \includegraphics{figures/khan.jpeg} & \includegraphics{figures/udacity.png} & \includegraphics{figures/coursera-production-result.jpg} & \includegraphics{figures/slide-person.jpg}\\
instructor, audience animation $\boxempty$ & instructor blackboard $\varbigcirc$ & pentip blackboard $\meddiamond$ & hand blackboard $\medtriangleright$ & talking-head slides, pentip $\medtriangledown$ & talking-head slides $\boxdot$\\
\end{tabular}

\includegraphics[width=.8\linewidth]{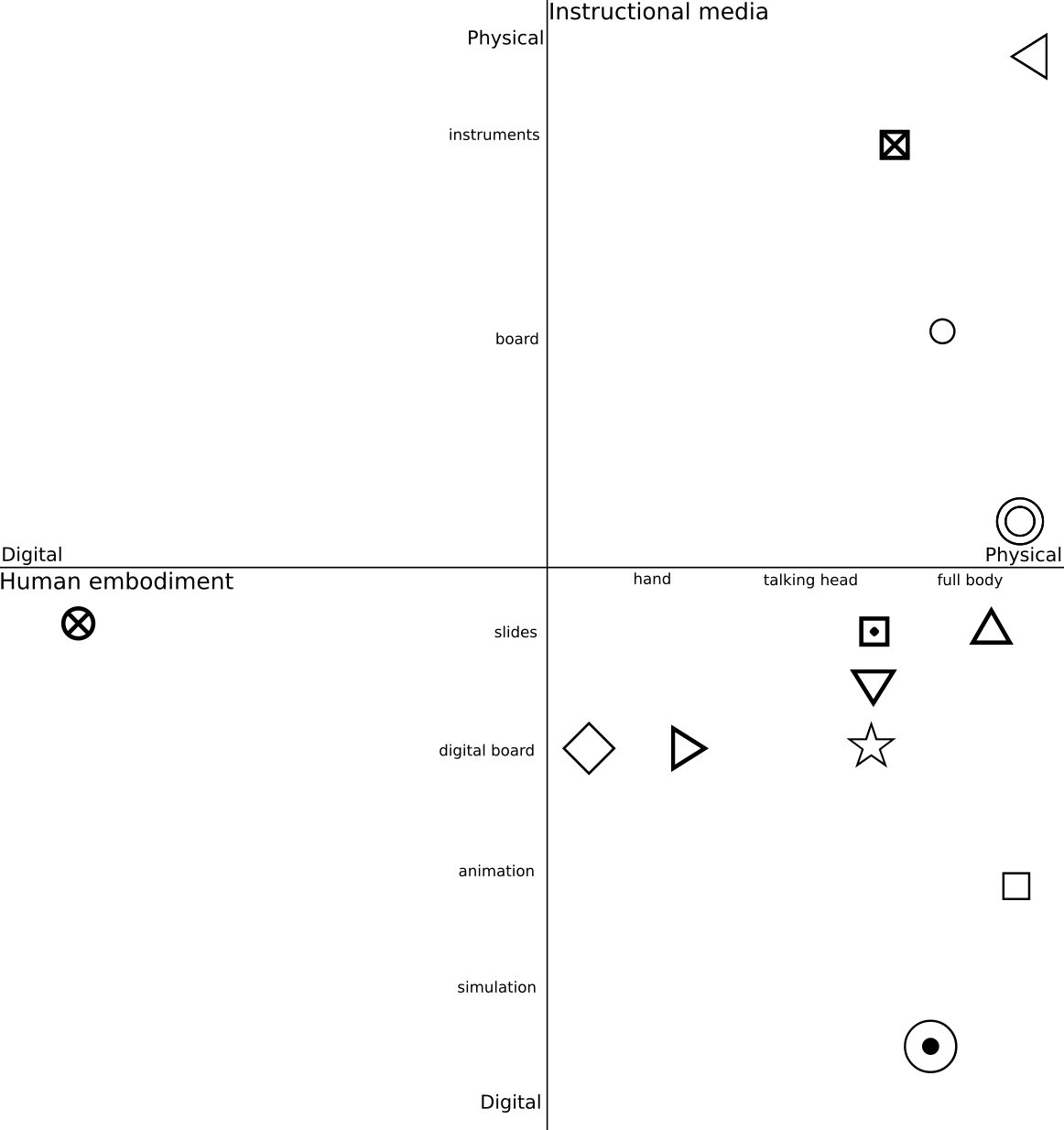}
\caption{Taxonomy of instructional video styles}

\centering
\fontfamily{ppl}\selectfont
\begin{tabular}{p{0.12\columnwidth}p{0.11\columnwidth}p{0.11\columnwidth}p{0.12\columnwidth}p{0.12\columnwidth}p{0.09\columnwidth}p{0.09\columnwidth}}
\includegraphics{figures/pip-lecture-style.jpg} & \includegraphics{figures/1975-mit-signal-processing-lab.jpg} & \includegraphics{figures/mit-architecture-final-review-committee.jpg} & \includegraphics{figures/guest-speaker.jpg} & \includegraphics{figures/shifman.png} & \includegraphics{figures/robot-slides.png} & \includegraphics{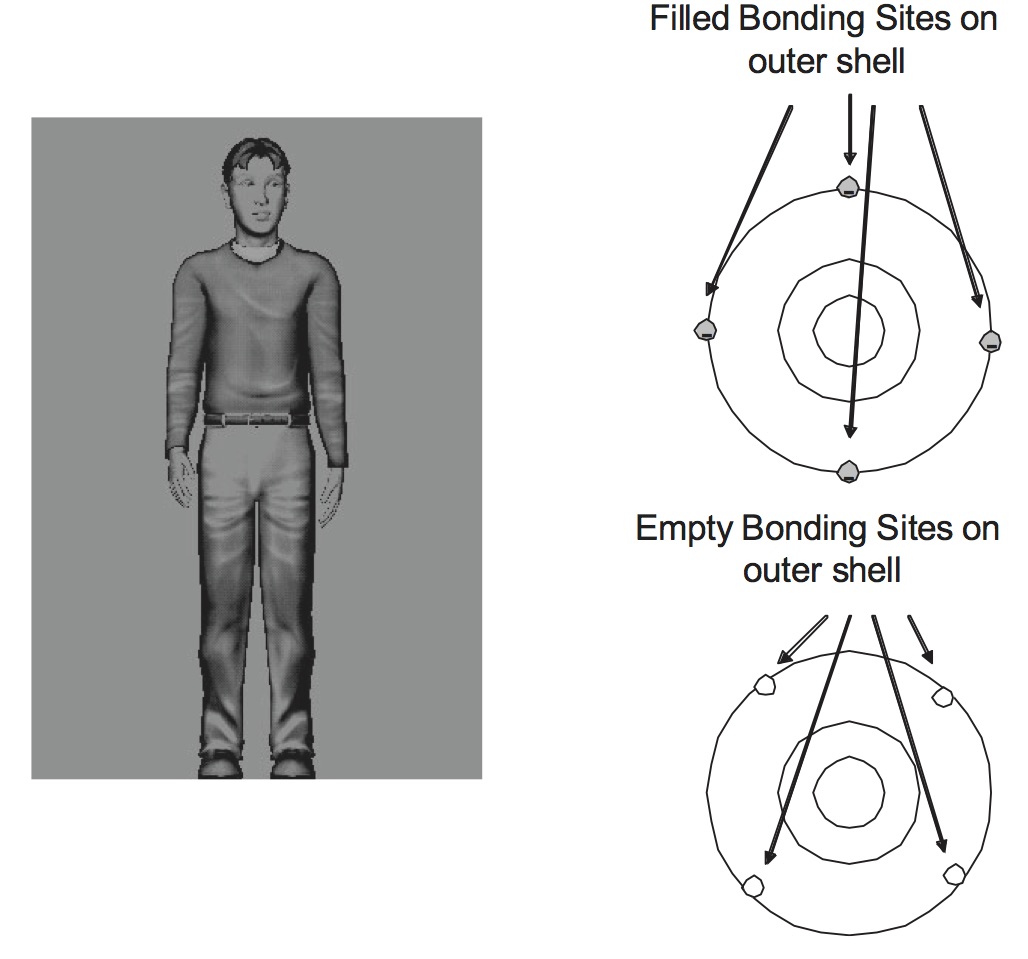} \\
talking-head pentip $\medwhitestar$ & talking-head instrument $\boxtimes$ & people instruments $\medtriangleleft$ & people no media $\ocirc$ & instructor coding $\varodot$ & robot slides  $\vartriangle$ & animated human slides $\otimes$ \\
\end{tabular}
\end{table*}

\section*{Discussion}

In comparison to previous related work, the proposed taxonomy: 1) describes current and potential new styles, 2) provides a granular spectrum of typologies, and 3) is complemented with a visual representation of video production styles. Most notably, Santos-Espino et al. (2016)\cite{santos2016speakers} have accurately identified the two main concepts (speaker-centric, board-centric), but they have only presented them as opposing conditions on one dimension. For example, they characterized existing instructional videos as speaker-centric, if the speaker is more important than the instructional media. In our view, the speakers and boards are not in a competition for the attention of the learner. Thus, the proposed taxonomy visually represents speakers and boards as two complementary (orthogonal) dimensions and, at the same time, it defines a wide continuum of presentation styles. Indeed, the survey of existing styles in research and practice has revealed that there is a more fine-grained use of human embodiment and instructional media. For example, speakers might be substituted with digital or artificial agents (animated characters, robots) and boards need not always be digital, but might be physical, too. Finally, the most important contribution of the suggested taxonomy is a comprehensive visual representation of existing and potential new presentation styles. In this way, the taxonomy of video lectures is more than a map of the current situation; it becomes a tool for navigating towards novel video lecture styles.

\begin{figure*}
\centering
\includegraphics[width=.4\linewidth]{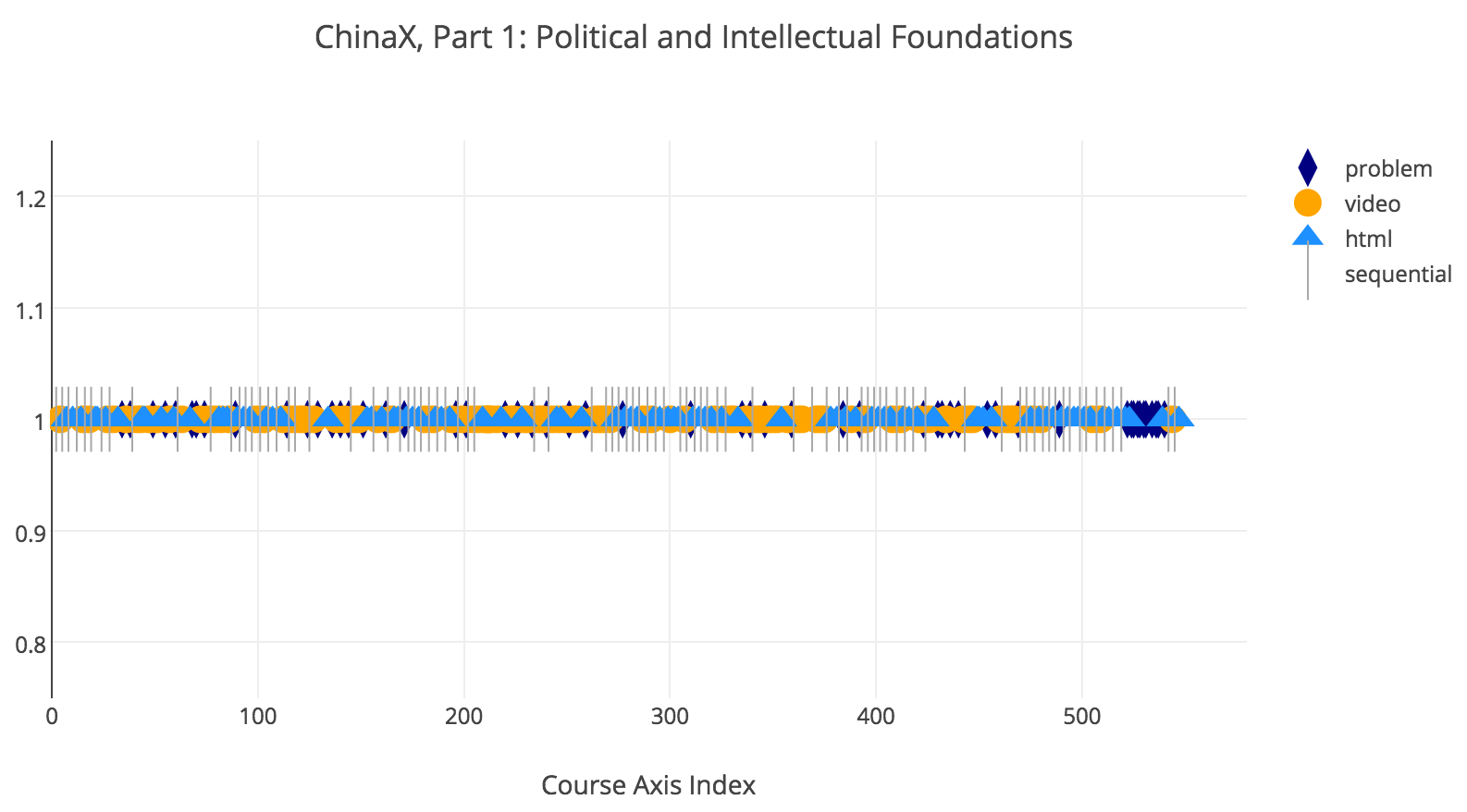}
\includegraphics[width=.3\linewidth]{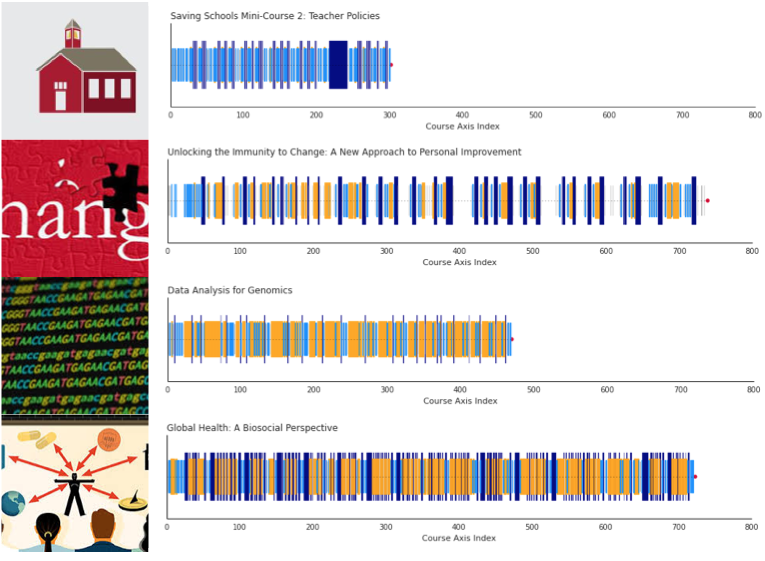}
\caption{visual
depiction (left) of a video-based course structure reveals that video is
just one component that needs to work with other equally important
components (e.g., problem sets, hypertext), while the actual mix of
these elements might differ significantly among courses (right)}
\end{figure*}

The visual representation of the instructional video taxonomy facilitates a systematic comparison between existing styles, as well as the design of new ones. It demonstrates that we are at an early phase in the development of instructional video styles, because most efforts just replicate the traditional classroom. Hopefully, there is a vast unexplored space that regards the employment of artificial representations for humans, such as robots and animated characters. In particular, it is worth exploring the combination of artificial characters with digital media, which might be facilitated by video-game development toolkits. In this way, digital characters might appear to manipulate digital instructional media in the third quadrant. Moreover, there are opportunities in the employment of augmented reality technologies, which bridge physical instructional media with artificial characters. For example, there are TV-studio technologies that enable the tracking of physical objects and enable their interaction with artificial entities (objects or characters). In this way, digital characters might appear to manipulate physical objects in the 4th quadrant.

A taxonomy of instructional video would not be complete unless we regarded the broader instructional framework. There are two main approaches to instructional video in education, which define a spectrum of options within them. Video lectures have been used as a substitute of classroom teaching in distance education, or as a complementary instructional tool in flipped-classrooms. For example, a video lecture prepared for distance education of adults assumes that the learner is going to have a minimal contact with the instructor and his peers. On the other hand, a video lecture prepared for K12 students, who attend school, assumes that the learner is going to employ the lecture as an instructional medium for home study. Therefore, the presentation style and the instructional design of video lectures might be influenced by the target group and the instructional framework. Moreover, according to Anderson and Dron (2010)\cite{anderson2011three} educational pedagogy could be classified in three generations: cognitive-behaviorist, social constructivist, and connectivist. The contemporary instructional video seems to be at the former stage with videos created by teachers and distributed to learners. In the future, video might be increasingly employed for peer-to-peer communication, or remixed and shared between learners and teachers. Therefore, further research could provide a taxonomy of video styles according to the pedagogical approach.

We did not consider neither the complete instructional design, nor the interaction design aspect of video lectures, but we focused on the visual organization of the video content. The instructional video is a major pillar in pedagogical design, but it is usually complemented with additional types of material (Seaton 2016)\cite{harvardx2016mooc}, so the selection of a presentation style should take a holistic view that considers the type of the course and the needs of the learners. For example, the Udacity video lectures are much more than video recordings of a teacher and instructional media, because they are highly structured in terms of learning design and provide the respective user interface that facilitates navigation through video and quiz content. The same practice is also followed by Coursera, but, the video segmentation seems to be more sparse than the one employed by the Udacity system. In addition to a video, there are more instructional materials, such as problem sets, hypertext pages, as well as discussion boards. Thus, further research should evaluate the effectiveness of presentation styles, in the context of particular pedagogical frameworks.

\section*{Conclusion}

In summary, there are some interesting patterns across the evolution and the production style of video lectures. First, video lectures have started as simple recordings of hour-long lectures and have gradually evolved into comprehensive one-minute clips of highly legible and elaborate tablet writing. In particular, there has been an increasing use of technology to manipulate the video recording of the teacher and of the instructional media. Most notably, there is wide variability of human embodiment in the final video, from groups of people, to robots, and digital avatars. In this way, human embodiment and instructional media have been two complementary dimensions in the proposed taxonomy that defines a highly granular two-dimensional space of existing and potential new presentation styles.

\marginnote{\emph{Media are mere vehicles that deliver instruction but do not influence
student achievement any more than the truck that delivers our groceries
causes changes in our nutrition, Clark (1983)}\cite{clark1983reconsidering}}

Besides the theoretical contribution (disambiguation of terms, granular two-dimensional taxonomy), the proposed taxonomy might facilitate the selection of a video lecture style, or it might encourage the production of novel ones. For example, a teacher might discover that the screencasting of slides might be enhanced with a drawing board, or a talking head video-feed, which add some extra personality to the final composite video lecture. Moreover, a more ambitious educator with access to studio equipment and with video production skills might discover that there is a vast unexplored space at the 3rd and especially in the 4th quadrant of the taxonomy. For example, a possible presentation style at the 4th quadrant might combine digital avatars that operate on physical instrumentation, which is a common special effect in movies. Therefore, in addition to some serious skills and equipment, we also need strong imagination, creativity, and experimentation in order to explore new presentation styles. Although the proposed taxonomy might be a necessary condition, it is certainly not a sufficient one for preparing successful video lecture, because an effective instructional video is a very complicated topic that also depends on pedagogy and production tools.

We expect that the experimentation with presentation styles will continue along a path similar to the experimentation of other linear audiovisual content, such as TV and radio. Indeed, the first TV shows were just radio shows with a static image (Bolter and Grusin 2000)\cite{bolter2000remediation}, but eventually, the TV format has evolved towards many novel directions. Similarly, we expect that instructional video is going to evolve away from simple classroom recording towards new presentation styles. In the race towards novel presentation styles, a few content providers might be able to afford expensive and elaborate production styles, which might result in high-quality audio-visual appeal and might raise the bar of what is expected by learners. Although the tools of video production have been democratized with inexpensive high-definition cameras and accessible post-processing at the home computer, there is still a big difference between an expensive Hollywood-studio movie and a low-budget independent production. At the same time, we expect that the mainstream instructional video might not be to the liking of everyone. Just like in movies, although production quality is important, there is always an audience for low-budget productions, which have to focus on other aspects, such as originality, narrative, and creative presentation style.

\bibliography{instructional-video-taxonomy}
\bibliographystyle{plainnat}

\end{document}